\documentclass[12pt]{article}

\catcode`\@=11
\@addtoreset{equation}{section}

\global\arraycolsep=2pt
\usepackage{amsbsy,amssymb,latexsym,amsfonts,amsmath}
\usepackage{graphicx,color}

\allowdisplaybreaks

\begin{document}
\begin{flushright}
\parbox{4.2cm}
{}
\end{flushright}

\vspace*{0.7cm}

\begin{center}
{\Large \bf 
Higher derivative corrections in holographic Zamolodchikov-Polchinski theorem}
\vspace*{2.0cm}\\
{Yu Nakayama}
\end{center}
\vspace*{-0.2cm}
\begin{center}
{\it California Institute of Technology, Pasadena, CA 91125, USA}\footnote{On leave from Berkeley Center for Theoretical Physics at University of California, Berkeley.}
\vspace{3.8cm}
\end{center}

\begin{abstract} 
We study higher derivative corrections in holographic dual of Zamolodchikov-Polchinski theorem that states the equivalence between scale invariance and conformal invariance in unitary $d$-dimensional Poincar\'e invariant field theories. From the dual holographic perspective, we find that a sufficient condition to show the holographic theorem is the generalized strict null energy condition of the matter sector in effective $(d+1)$-dimensional gravitational theory. The same condition has appeared in the holographic dual of the ``c-theorem" and our theorem suggests a deep connection between the two, which was manifested in two-dimensional field theoretic proof of the both.
\end{abstract}

\thispagestyle{empty} 

\setcounter{page}{0}

\newpage

\section{Introduction} 
Is scale invariance equivalent to conformal invariance? The recurrent theme of this paper  is one of the important unsolved problems in the quantum field theory and our modern physics.  The superficial answer to the question is ``No". The conformal invariance is more than the mere scale invariance. Nevertheless, in the modern quantum field theories, we cherish the belief that the scale invariance and the conformal invariance are equivalent under some mild assumptions. Indeed, we do not know any examples of scale invariant but non-conformal field theories once we assume unitarity, Poincar\'e invariance and discreteness of the spectrum. 

Some of our understandings of the modern physics crucially depend on the equivalence between the scale invariance and conformal invariance. For instance, in the world-sheet formulation of the string theory, we claim that the target space equation of motion can be obtained by imposing the vanishing of the beta function of the underlying two-dimensional world-sheet field theories, but the vanishing of the beta function only indicates scale invariance, and does not automatically imply conformal invariance. It is the conformal invariance that is necessary for the consistency of the world-sheet string theory. Can we imagine an inconsistent string background that satisfies  ``the string equation of motion" just because of the lack of conformal invariance? How can it be?

Another important example arises in the condensed matter physics. In order to classify the possible critical phenomena, we use the conformal field theory techniques, in particular its classification.  The strategy to classify all the critical phenomena by classifying the possible conformal field theories is quite successful in $(1+1)$ dimension (or equivalently two-dimensional classical system). However, this classification entirely relies on our belief that the scale invariance is equivalent to the conformal invariance. Is it really the case that the critical phenomena must always show the conformal invariance while its definition only requires the scale invariance?  Is there any deep physics behind the distinction between scale invariance and conformal invariance?

From the field theory perspective, the only known result concerning this problem is the theorem proved by Zamolodchikov and Polchinski in $(1+1)$ dimension \cite{Zamolodchikov:1986gt}\cite{Polchinski:1987dy} (see also \cite{Mack1}). They showed that in $(1+1)$ dimension, the scale invariant field theory is always conformally invariant under the assumption that the theory is unitary, Poincar\'e invariant, and it has a discrete spectrum. In higher dimensions, however, there is no proof or counterexamples of the equivalence between the scale invariant but non-conformal field theories.

The holographic approach to the problem was initiated in \cite{Nakayama:2009qu}\cite{Nakayama:2009fe}, where it was shown that the null energy condition for the matter energy momentum tensor guarantees the enhancement of the symmetry of the gravitational field configuration from the dilatation group with the Poincar\'e symmetry to the full conformal group (AdS isometry). The discussion there was restricted to the Einstein gravity, and the aim of this paper is to generalize the discussion with higher derivative corrections.

Thus, in this paper, we study higher derivative corrections in holographic dual of the Zamolodchikov-Polchinski theorem that states the equivalence between scale invariance and conformal invariance in unitary Poincar\'e invariant field theories. From the dual holographic perspective, we find that a sufficient condition is the generalized strict null energy condition of the matter sector in effective $(d+1)$ dimensional gravitational theory. The same condition has appeared in the holographic dual of the ``c-theorem" (see \cite{Girardello:1998pd}\cite{Freedman:1999gp} and its higher derivative generalization in \cite{Myers:2010xs}\footnote{See also the cosmological discussions in \cite{Strominger:2001gp} and \cite{Sinha:2010pm}.}) and our theorem suggests a deep connection between the two, which is manifested in $(1+1)$ dimensional field theoretic proof of the both \cite{Zamolodchikov:1986gt}\cite{Polchinski:1987dy}.
\section{Holographic Zamolodchikov-Polchinski theorem}
In $(1+1)$ dimension, the equivalence between the scale invariance and the conformal invariance was proved by Zamolodchikov and Polchinski. The theorem holds under the assumptions
\begin{itemize}
	\item the theory is unitary
	\item the theory is Poincar\'e invariant, and
	\item the theory has a discrete spectrum.
\end{itemize}
It has been conjectured that the theorem is true in higher dimensions as well, but there is no proof or counter-example in higher dimensions than two (see. e.g. \cite{Dorigoni:2009ra} and references therein).

In this section, we would like to establish the holographic Zamolodchikov-Polchinski theorem \cite{Nakayama:2009fe}. The holographic theorem applies not only in bulk $(2+1)$ dimension but in any dimension.

{\bf Holographic dual of Zamolodchikov-Polchinski theorem}

{\it When the geometry as well as all the other field configurations are invariant under the dilatation as well as $d$-dimensional Poincar\'e group, the symmetry is enhanced to the AdS$_{d+1}$ group, which guarantees the conformal invariance of the dual field theory.}\footnote{The algebra for the isometry of the Poincar\'e group with dilatation as well as the full conformal group ($\simeq$ AdS group) are summarized in Appendix.}

Note that our assumption is that the symmetry is realized as an isometry of the gravitational field configuration. This is not necessarily required: the Virasoro symmetry in AdS$_3$ or extremal Kerr black hole \cite{Guica:2008mu} is not realized by the isometry but as an asymptotic symmetry group. In this paper, we focus on the simpler case when the full symmetry is realized as an isometry.

In this section, we would like to review the proof of this theorem in effective Einstein gravity coupled with the matter that satisfies the strict null energy condition. In the next section, we would like to pursue the generalization of the holographic theorem in higher derivative gravitational theories.

\subsection{Pure gravity}
We would like to begin with the search for deformations of the AdS space with scale invariance but without special conformal invariance. In the effective gravity approach, a $d$-dimensional conformal field theory is dual to a gravitational theory on the $(d+1)$-dimensional AdS space whose metric is given by
\begin{align}
ds^2 = \frac{dz^2 + \eta_{\mu\nu} dx^\mu dx^\nu}{z^2} \ .
\end{align}
We exclusively use the Poincar\'e metric in the following because we would like to keep the invariance under the Poincar\'e group manifest.

The AdS space is obviously invariant under the Poincar\'e group acting on the $x^\mu$ plane, and the dilatation:
\begin{align}
x^{\mu} \to \lambda x^{\mu} \ , \ \  z \to \lambda z \  \label{scale}
\end{align}
In addition, it is invariant under the special conformal transformation.
\begin{align}
\delta x^\mu = 2(\epsilon^{\nu}x_\nu)x^\mu - (z^2 + x^\nu x_\nu)\epsilon^\mu \ , \ \ \delta z = 2(\epsilon^{\nu}x_\nu)z  \label{cnft}
\end{align}
as an isometry. The AdS space is a maximally symmetric solution of the Einstein equation with a negative cosmological constant. For later purposes, we mention that a $(d+1)$-dimensional manifold is maximally symmetric if and only if the Riemann tensor obeys the relation:
\begin{align}
R_{abcd} = \frac{R}{d(d+1)} (g_{ac}g_{bd} - g_{ad}g_{bc}) \ .
\end{align}

The question we would like to ask here is: Is there any scale invariant but non-conformal deformation of the AdS geometry?
It is easy to convince ourselves that there is no such a deformation at all. Indeed, we can state the following proposition:

{\bf Proposition } 

{\it $(d+1)$ dimensional pseudo Riemannian manifold whose isometry contains $d$-dimensional Poincar\'e group together with the dilatation must be (the Poincar\'e patch of) AdS$_{d+1}$ space  with the enhanced symmetry of the conformal group.}

The argument is simple. As in the Bianchi classification of pseudo Riemannian geometries, the invariance under the $d$-dimensional Poincar\'e group demands that the metric take the following form:
\begin{align}
ds^2 = f(z) dz^2 + g(z) \eta_{\mu\nu} dx^\mu dx^\nu \ .
\end{align}
If we further demand the dilatation invariance under $z \to \lambda z$, $x^\mu \to \lambda x^\mu$, the warp factor is uniquely determined as $f(z) = g(z) = z^{-2}$ up to diffeomorphism and an overall factor. The consequent metric is nothing but the Poincar\'e patch of the AdS space.

 The proposition simply indicates that the scale invariant but non-conformal geometry is impossible within the pure gravity whose field theory dual contains only the energy-momentum tensor. Of course, this is expected from the field theory side because the only possible non-trivial scalar deformation that is scale invariant and can be constructed out of the energy momentum tensor alone would be $\int d^dx T^{\mu}_{\ \mu}$, but it simply vanishes due to the scale invariance.\footnote{Note that $T^{\mu}_{\ \mu}$ does not necessarily vanish, but its integral does in scale invariant field theories.}

\subsection{With matters satisfying null energy condition}

In section 2.1, we have seen that the scale invariant but non-conformal geometry does not exist in the effective gravity approach. The field theory interpretation was obvious and it is now clear that we have to introduce non-trivial matters in the gravity side to pursue possibilities for scale invariant but non-conformal field configuration.

A typical example of such field configurations is a condensation of the form field. For instance, the one-form field that has a profile
\begin{align}
A = A_a dx^a = a \frac{dz}{z} \label{av}
\end{align}
is invariant under the dilatation and the Poincar\'e group but it is not invariant under the special conformal transformation. Similarly, the $d$-form field
\begin{align}
B = b \frac{dx^0\wedge \cdots \wedge dx^{d-1}}{z^d} \label{bv}
\end{align}
is invariant under the dilatation and the Poincar\'e group but it is not invariant under the special conformal transformation.
We note that we have assumed that the form field introduced here does not possess any gauge invariance so that non-zero value of $a$ and $b$ physically breaks the special conformal invariance.

The only non-trivial form field configuration that is invariant under the full AdS isometry dual to the conformal group is the $d+1$-form field constantly proportional to the volume form:
\begin{align}
V = v \frac{dz \wedge dx^0\wedge \cdots \wedge dx^{d-1}}{z^{d+1}} \ ,
\end{align}
which is dual to a cosmological constant.

The gravitational analogue of the Zamolodchikov-Polchinski theorem must dictate  $a = b = 0$. Since such a scale invariant but non-conformal field configuration exists, the real question is whether such a condensation can or cannot occur once we imposed the equations of motion. The aim of this subsection is to explain how this can be realized by imposing the ``healthy" bulk equations of motion.

Before we proceed, as we have noted that we need three assumptions in the original Zamolodchikov-Polchinski theorem, we naturally expect that we need some conditions in the dual gravity side to establish the theorem if any. Out of the three assumptions,  the Poincar\'e invariance is manifest from our ansatz and the discreteness of the spectrum is also guaranteed if we introduce a finite number of bulk fields. The last non-trivial assumption is the unitarity of the boundary theory. The corresponding assumption that we introduce in order to derive the gravity analogue of the Zamolodchikov-Polchinski theorem is the so-called null energy condition. 

In general relativity there are several notions of energy condition that guarantee ``good behavior" of the energy-momentum tensor. The null energy condition is the weakest form of the energy condition, and it plays a significant role in the thermodynamics of black holes and the stability of the space-time.

The null energy condition demands that the right hand side of the particular combination of the Einstein equation is semi-positive definite. We know that the scale invariant geometry is necessarily AdS space, and the null energy condition means\begin{align}
R_{zz} + R_{00} \ge 0 \ ,
\end{align}
where the inequality is useful when we replace the Ricci tensor with the matter energy momentum tensor by the Einstein equation. 
\begin{align}
R_{ab} - \frac{R}{2}g_{ab} = \kappa T_{ab} 
\end{align}
or
\begin{align}
R_{ab} = \kappa\left(T_{ab} - \frac{1}{(d+1)-2}T g_{ab} \right) 
\end{align}

The important observation is that after replacing the Ricci tensor by the trace reversed Einstein equation, the combination appears in the null energy condition is nothing but the flux that would violate the special conformal invariance. In the above examples of form field, it is given by
\begin{align}
z^2(R_{zz} + R_{00}) = C_1(a,b)a^2 + C_2(a,b)b^2 \ , \label{nec1}
\end{align}
where the coefficient $C_1$ and $C_2$ are semi-positive definite when the theory satisfies the null energy condition.

For instance, let us consider the Proca action with a certain forth order interaction $\frac{1}{4}F_{ab} F^{ab} + m^2 A_a A^a +\lambda (A_a A^a)^2$, which can be obtained by a (higher derivative) Higgs model in the  unitary gauge. With the vector condensation \eqref{av}, the right hand side of \eqref{nec1} is given by $ (m^2 + 2\lambda a^2) a^2$. The stability of the action, which amounts to the null energy condition, demands $m^2 \ge 0$ and $\lambda \ge 0$ so that $C_1 \ge 0$ in \eqref{nec1}.

However, we know that the left hand side of \eqref{nec1} is zero because we know that the metric is the AdS metric when the gravitational field configuration is scale invariant. Then we immediately conclude that $a^2 = b^2 =0$ so that the field configuration is not only scale invariant but also conformally invariant: it has a full AdS isometry. When $m^2 = \lambda = 0$, $a \neq 0$ may be allowed, but then the massless Proca equation cannot eliminate the unitarity violating transverse degree of freedom and the theory is sick. To exclude the possibility from our discussion, we will slightly generalize the null energy condition to the so-called ``strict null energy condition".

The strict null energy condition states that the equality of the null energy condition is satisfied if and only if the field contributing to the energy momentum tensor takes the trivial value \cite{No}. More precisely, we demand that if there exists any null-vector that makes $T_{ab} k^a k^b =0$, then the field configuration must be invariant under all the isometry transformation of the space-time.  Note that the null energy-condition itself cannot exclude the sick non-unitary matter configuration of the 1-form condensation with $m^2 = \lambda = 0$ mentioned in the previous paragraph, but the strict null energy condition does.

We now would like to generalize the discussion to arbitrary higher spin tensor fields $\mathcal{T}$. For this, we note that that the requirement of the scale invariance is that they are invariant under the Lie derivative with respect to the isometry corresponding to the scale transformation
\begin{align}
\mathcal{L}_{v_d} \mathcal{T} = 0 \ ,
\end{align}
while they are not invariant under the Lie derivative with respect to the isometry corresponding to the special conformal transformation
\begin{align}
\mathcal{L}_{v_s} \mathcal{T} = 0 \ .
\end{align}

By inspecting the tensor structure, such tensor fields generically contribute to the right hand side of the null energy condition by using the Einstein equation as in \eqref{nec1}.\footnote{Schematically, the energy momentum tensor of the tensor matter has the structure $g_{ab} H(\mathcal{T}) + \mathcal{T}_{a c \cdots}\mathcal{T}_{b}^{\ d \cdots} G^{c \dots}_{d \dots}(\mathcal{T})$, and the first term gives zero contribution to the null energy-condition in AdS space, but the second term gives the non-zero contribution. The claim here is that for the higher theory to be ``healthy", $G^{c \dots}_{d \dots}$ must be constrained so as to satisfy the (strict) null energy condition.} 
However, the left hand side vanishes because the geometry is AdS space.
Here, we will use the notion of the strict null energy condition introduced above. When the matter fields satisfy the strict null energy condition, we can conclude that the condensation of the higher tensor fields that will break the conformal invariance cannot occur, and hence it proves the holographic Zamolodchikov-Polchinski theorem.

In the context of the black hole holography, the significance of the null energy condition is well-appreciated because it gives the sufficient condition for the  area non-decreasing theorem of the black hole horizon, which has the interpretation of the non-decreasing property of the black hole entropy. The strict null energy condtion roughly states that when the entropy stays the same, nothing non-trivial happens. In other words, ``zero-energy state" cannot contain any information. We will come back to the holographic meaning of the (strict) null energy condition in our setup in section 4.

\section{Generalized null energy-condition and higher derivative corrections}

The main goal of this paper is to generalize the discussion in the previous section to more general theories of gravity with higher derivative corrections. From the holographic viewpoint, the higher derivative corrections amount to including $1/N$ corrections to the dual field theory. It would be of importance to understand the condition for the holographic Zamolodchikov-Polchinski theorem to hold. In particular, in higher dimension than two, there is no known field theory result, so the holographic approach is all the more relevant.

Our claim is that a sufficient condition that the holographic Zamolodchikov-Polchinski theorem can be shown in the higher derivative gravitational theory is the generalized strict null energy condition. We note that the same generalized null energy condition was the basis of the holographic ``c-theorem" in the higher derivative gravity from the effective field theory approach \cite{Freedman:1999gp}\cite{Myers:2010xs}.

Even with the higher derivative corrections, the proposition discussed in section 2.1  still applies: the scale invariant geometry with the $d$-dimensional Poincar\'e invariance in $(d+1)$-dimensional space is always AdS$_{d+1}$. Therefore, the violation of the conformal invariance only comes from the non-trivial matter configuration as in section 2.2. As discussed there, the requirement of the scale invariance is that they are invariant under the Lie derivative with respect to the isometry corresponding to the scale transformation
\begin{align}
\mathcal{L}_{v_d} \mathcal{T} = 0 \ ,
\end{align}
while they are not invariant under the Lie derivative with respect to the isometry corresponding to the special conformal transformation
\begin{align}
\mathcal{L}_{v_s} \mathcal{T} = 0 \ .
\end{align}
By inspecting the tensor structure, such tensor fields generically contribute to the right hand side of the generalized null energy condition.

The generalized null energy condition is formulated in the following way. In higher derivative gravity, the equation of motion is frame-dependent, so let us consider the metric equation of motion in a certain frame. The equation of motion for the metric can be written in the following form:
\begin{align}
\tilde{G}_{ab} = \tilde{T}_{ab} \ , \label{gein}
\end{align}
where $\tilde{G}_{ab}$ is the tensor made from metric, curvature, and their derivatives but with no contribution from the matter fields. $\tilde{T}_{ab}$ is the generalized energy momentum tensor that is made from matter fields (possibly with gravitational higher derivative corrections). The generalized null energy condition means that in a certain frame, the matter energy momentum tensor satisfies the inequality:
\begin{align}
k^a k^b \tilde{T}_{ab} \ge 0 \ , \label{nuls}
\end{align}
for any null vector $k^a$ (i.e. $k^a k_a= 0$). When the equality is only satisfied for a trivial field configuration, the energy condition acquires an adjective ``strict". In particular, the strict condition in our case implies that all the field configuration must be invariant under the Lie derivative with respect to the AdS isometry.

Let us focus on the frame where the strict null energy condition \eqref{nuls} holds. In that frame,\footnote{Actually, the statement here must be true in {\it any} frame when all the field configuration is invariant under the dilatation group. When all the field configuration is invariant under the Lie derivative corresponding to the dilatation, the redefinition of the metric does not affect the isometry group.} as discussed in the above, the scale invariant geometry is the AdS space. The AdS space is the maximally symmetric space, so every curvature invariant and its derivative must be rewritten by the metric tensor itself. In particular, in the generalized ``Einstein equation" \eqref{gein}, the left hand side must be proportional to the metric tensor $\tilde{G}_{ab} = c g_{ab}$. Now one can choose a particular null vector $k^a$ (i.e. $k^0=k^z = 1$) and study the strict null energy condition:
\begin{align} 
\tilde{T}_{00} + \tilde{T}_{zz} = \tilde{G}_{00} + \tilde{G}_{zz} = c\left(g_{00} + g_{zz}\right) = 0  \ . 
\end{align} 
The first equality was the generalized Einstein equation, and the next (and the last) equality came from the property of the maximally symmetric space.
 
 Since the left hand side of the above equality is what we have in the generalized null energy condition, the strict version of the generalized null energy condition demands that the matter field configuration must be trivial. All the matter fields must be invariant under the full AdS isometry. In particular, the matter configuration that would violate the conformal invariance such as \eqref{av} or \eqref{bv} is forbidden by the strict null energy condition.

For instance, let us consider the Gauss-Bonnet-Einstein gravity in $(1+4)$-dimension:
\begin{align}
S = \frac{1}{2\kappa} \int d^5 x \sqrt{-g}( R + \alpha L -2\Lambda) + S_{\mathrm{matter}} \ ,
\end{align}
where the Gauss-Bonnet term is $L = R_{abcd}R^{abcd} - 4R_{ab}R^{ab} + R^2$. The metric equation of motion is given by
\begin{align}
G_{ab} + \Lambda g_{ab} + \alpha\left( 2R_{almn}R_b^{lmn} - 4R_{ambn}R^{mn} - 4R_{am}R_{b}^m + 2RR_{ab} - \frac{1}{2} g_{ab} L \right) = \kappa T_{ab} \ . \label{GBE}
\end{align}
We evaluate the left hand side of \eqref{GBE} in the AdS space, and it immediately gives us the equation $T_{00} + T_{zz} = 0$. The generalized null energy condition applied for $T_{ab}$ demands that the non-trivial field condensation that violates the AdS isometry is impossible so that the Zamolodchikov-Polchinski theorem holds.

As an example of the  matter action, let us take the massive vector model:
\begin{align}
S_{\mathrm{matter}} = \int d^5 x \sqrt{-g} \left(\frac{1}{4} F_{ab}F^{ab} + V(A_a A^a) \right) \ , 
\end{align}
where $F_{ab} = \partial_a A_b - \partial_b A_a$ as usual. For the scale invariant but non-conformal field configuration $A_a dx^a = \frac{c dz}{z}$, the energy-momentum tensor is proportional to $V'(A_a A^a) g_{ab}$. When the equation of motion $V'(A_a A^a) = 0$ admits a non-trivial solution $c\neq 0$, the holographic Zamolodchikov-Polchinski theorem is violated, but then at the same time, the (strict) null energy condition is violated as well (and vice versa).

As in the Einstein case, one can find counterexamples of holographic Zamolodchikov-Polchinski theorem by introducing the matter that violates the (generalized) null energy condition (see the above example). Another typical example is the ghost-condensation as shown in \cite{Nakayama:2009fe}.  In particular, in the Euclidean gravity, there is no notion of energy condition so that we expect physical example of violation of the Zamolodchikov-Polchinski  theorem.\footnote{Note that in the Euclidean signature, the unitarity is replaced by the reflection positivity, and the latter is not always required. As a consequence, there exist scale invariant but non-conformal field theories in Euclidean signature with physical significance \cite{Nakayama:2010ye}\cite{Nak}.}

\section{Discussion}
In this paper, we have studied higher derivative corrections in the holographic approach to the Zamolodchikov-Polchinski theorem. Our result shows that as long as the matter field satisfies the generalized strict null energy condition, the scale invariant $d+1$-dimensional gravitational field configuration has an enhanced AdS symmetry, which indicates the conformal invariance translated in the dual field theory.

Does it mean that the gravitational sector can possess an arbitrary higher derivative effective action, which cannot be regulated from the dual field theory? The answer is not always. To answer the question, it is imperative to recall that our analysis is entirely based on the assumption that the gravitational field configuration has $d$-dimensional Poincar\'e invariance. As an immediate consequence, the effect of compactification is all neglected, or rather it has been implemented implicitly in the behavior of the matter action.

For instance, even when the higher derivative gravity has a non-unitary spectrum or non-unitary field theory dual, our theorem still applies. In such cases, can we say anything about the (in)consistency of the theory from the holographic Zamolodchikov-Polchinski theorem as first advocated in \cite{Nakayama:2009qu}? The key point is to study the lower-dimensional compactified theory to apply the holographic Zamolodchikov-Polchinski theorem. Suppose we compactify an extra dimension, and consider, say $(d-1)$-dimensional Poincar\'e invariant field configuration with an extra dilatation isometry. The Kaluza-Klein vector field typically acquires a non-unitary ghost-like (higher derivative) kinetic term inherited from the original gravitational kinetic terms. The Kaluza-Klein fields now break the null energy condition, and the theorem discussed in the paper does not apply. In this way, the non-unitary kinetic term in the  gravitational sector might be incompatible with the holographic Zamolodchikov-Polchinski theorem after compactification.

As proposed in \cite{Nakayama:2009qu}, it would be of great importance to study the constraint on the effective field theories from the holographic viewpoint. The compactification effect within the string theory and M-theory in our context was studied in \cite{Nakayama:2009fe} at the tree level. It would be interesting to perform the more systematic analysis with the higher derivative corrections to discuss the consistency of the higher derivative gravitational corrections with compactification.

Finally, we would like to mention the relation between our theorem and the holographic ``c-theorem" with the higher derivative corrections. The both theorems have field theoretic proofs in two-dimension with essentially the same additional assumptions such as unitarity and Poincar\'e invariance.
Without the higher derivative corrections, the holographic counterpart of the both theorem was proved by using the null energy condition \cite{Freedman:1999gp}. It has been recently discussed \cite{Myers:2010xs} that the the generalized null energy condition is a sufficient condition to prove the holographic ``c-theorem" with the higher derivative corrections. 
The discussion in this paper suggests a further link between the two theorems even in higher dimension. 

Indeed, when we compare our discussion with the holographic c-theorem, in particular in $(1+1)$-dimensional dual case where the proof of the both is available, we are able to understand how the strict null energy condition is essential in our argument as well as theirs. The holographic c-theorem roughly tells that the change of the central charge with respect to the scale $\mu$ is bounded by the null-energy condition:
\begin{align}
 \frac{d c}{d \log \mu} = T_{00} + T_{zz}  \ge 0 \ .
\end{align}
In $(1+1)$ dimension, it was shown \cite{Zamolodchikov:1986gt} that the left hand side is the two-point function of the trace of the energy-momentum tensor $\langle T(x) T(0) \rangle$, and it vanishes if and only if the theory is conformal (and the dual geometry has AdS isometry). The strict null energy condition guarantees this ``only if" part of the argument that comes from the unitarity of the $(1+1)$-dimensional field theory. Thus, the strict null energy condition is sufficient for the holographic c-theorem to be fully consistent with the field theory derivation in $(1+1)$ dimension, and the null energy-condition guarantees the unitarity of the dual field theory. Our independent gravitational derivation of the Polchinski-Zamolodchikov theorem using the strict null energy condition further strengthens this viewpoint.

\section*{Acknowledgements}
The work was supported in part by the National Science Foundation under Grant No.\ PHY05-55662 and the UC Berkeley Center for Theoretical Physics.

\appendix

\section{Conformal algebra}
A fundamental symmetry of quantum field theories is the Poincar\'e algebra:
\begin{align}
i[J^{\mu\nu},J^{\rho\sigma}] &=
\eta^{\nu\rho}J^{\mu\sigma}-\eta^{\mu\rho}J^{\nu\sigma} -
\eta^{\sigma\mu}J^{\rho\nu} + \eta^{\sigma\nu}J^{\rho\mu} \cr
i[P^\mu,J^{\rho\sigma}] &= \eta^{\mu\rho}P^{\sigma} -
\eta^{\mu\sigma} P^\rho \cr[P^\mu,P^\nu] &= 0 \ .
\end{align}
For massless scale invariant theory, one can augment this Poincar\'e
algebra by adding the dilatation operator $D$ as
\begin{align}
[P^\mu,D] &= i P^\mu \cr [J^{\mu\nu},D] &= 0  \ .
\end{align}
The maximally enhanced bosonic symmetry
of the space-time algebra for massless particles is given by the
conformal algebra (plus some internal symmetries):
\begin{align}
[K^\mu,D] &= -iK^\mu \cr [P^\mu,K^\nu] &= 2i\eta^{\mu\nu}D+
2iJ^{\mu\nu} \cr [K^\mu,K^\nu] &= 0 \cr [J^{\rho\sigma},K^\mu] &=
i\eta^{\mu\rho} K^\sigma - i\eta^{\mu\sigma} K^\rho \ ,
\end{align}
where $K^\mu$ generate special conformal transformation. As a group, the conformal group is isomorphic to the isometry of the AdS space.

\end{document}